\documentclass{Interspeech2024}

\usepackage{xcolor}
\usepackage{graphicx}
\usepackage{amssymb}
\usepackage{amsmath}
\usepackage{booktabs}
\usepackage{makecell}
\usepackage{pifont}
\usepackage{multirow}
\newcommand{\xmark}{\ding{55}}

\interspeechcameraready

\title{Transfer Learning from Whisper for Microscopic Intelligibility Prediction}

\name[]{Paul}{Best$^*$}
\name[]{Santiago}{Cuervo$^*$}
\name[]{Ricard}{Marxer}

\address{
  Université de Toulon, Aix Marseille Univ, CNRS, LIS, Toulon, France\thanks{$^*$Equal contribution.}}
\email{\{paul.best,santiago.cuervo,ricard.marxer\}@lis-lab.fr}


\begin{document}

\maketitle

\begin{abstract}  
    Macroscopic intelligibility models predict the expected human word-error-rate for a given speech-in-noise stimulus. In contrast, microscopic intelligibility models aim to make fine-grained predictions about listeners' perception, e.g. predicting phonetic or lexical responses. State-of-the-art macroscopic models use transfer learning from large scale deep learning models for speech processing, whereas such methods have rarely been used for microscopic modeling. In this paper, we study the use of transfer learning from \emph{Whisper}, a state-of-the-art deep learning model for automatic speech recognition, for microscopic intelligibility prediction at the level of lexical responses. Our method outperforms the considered baselines, even in a zero-shot setup, and yields a relative improvement of up to 66\% when fine-tuned to predict listeners' responses. Our results showcase the promise of large scale deep learning based methods for microscopic intelligibility prediction.
\end{abstract}

\section{Introduction}

Traditional speech intelligibility models measure how comprehensible speech is. Generally, they make predictions about the average number of words that may be heard correctly given a noisy speech stimulus. These models are referred to as \emph{macroscopic}, as their prediction target is an expected value aggregated across many words or sentences, and multiple listeners. Macroscopic intelligibility prediction has been of great interest for its importance in developing speech enhancement related  applications, such as hearing aids. 

In the last decade, a different paradigm for intelligibility modeling has emerged: \emph{microscopic intelligibility prediction} \cite{marxer2015framework, mandel13}. These models aim to predict human perception of speech at a higher level of detail than macroscopic models. For instance, such model might be expected to predict a confusion matrix characterising phonetic misperceptions, or to predict a listener’s lexical response to a specific noisy speech token. It is hoped that building these models will contribute to understanding the mechanisms underlying human speech perception, since they must reproduce more specific listener's behavior.


Throughout the last decade, we have also witnessed great advances in speech processing applications thanks to the popularization and scaling of deep learning methods. Training large-scale deep learning models from scratch on intelligibility prediction is usually not feasible due to the small amount of labeled data available. However, recent advances on representation learning showed that transfer learning from large-scale self-supervised or weakly-supervised pre-trained models is highly effective across a wide range of speech applications \cite{baevski2020wav2vec, chen2021wavlm,radford2023robust, gong23whisperat}. These methods yielded the state-of-the-art models for macroscopic intelligibility prediction \cite{tu22b_interspeech, cuervo2024speech}, but little work has been done on applying them to microscopic modeling. Moreover, there is evidence that the representations learned by large scale pre-trained models are predictive of human brain speech processing \cite{millet2022toward}, likely making them relevant to model speech perception with tasks such as microscopic intelligibility prediction.

In this paper, we explore the use of transfer-learning from a state-of-the-art large-scale deep learning model for automatic speech recognition (ASR) to microscopic intelligibility prediction. We focus on \emph{Whisper} \cite{radford2023robust}, an ASR model that approaches the accuracy and robustness of humans. We use it to tackle the task of predicting lexical responses, which is considered to be the most challenging on microscopic intelligibility modeling \cite{marxer2015framework}. Our method outperformed the considered baselines, even in a zero-shot setup. When fine-tuned to predict listener's responses, it yielded a relative improvement of 66\%. Overall, our results showcase the promise of transfer learning from large pre-trained models to microscopic intelligibility prediction.




\begin{figure*}
    \centering \vspace{-.5cm}
    \includegraphics[width=0.7\linewidth]{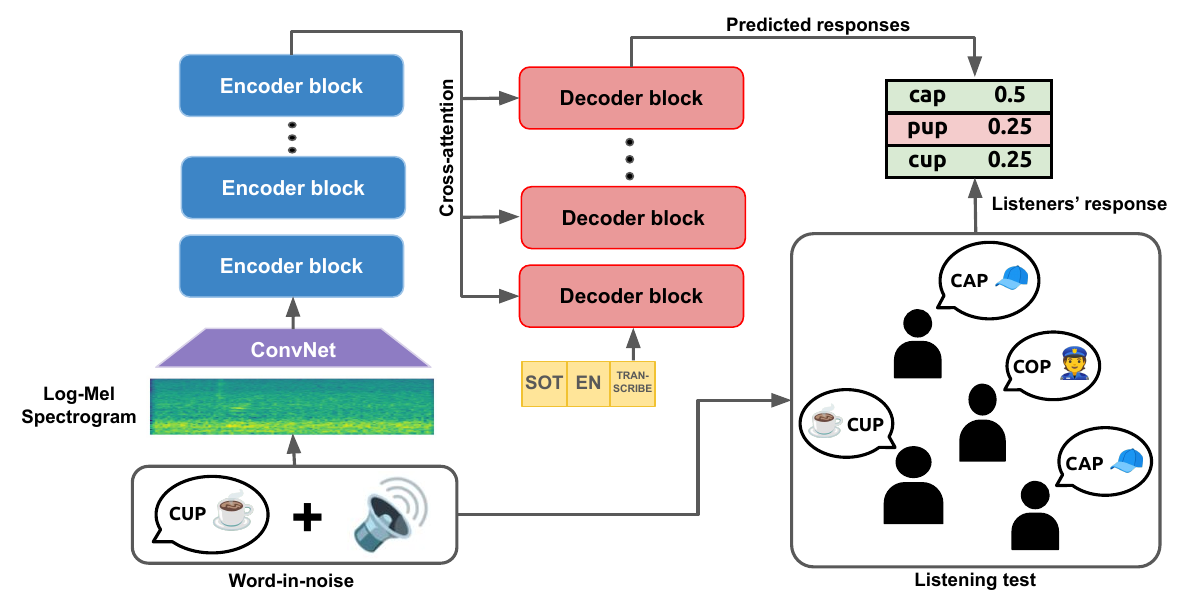}
    \caption{Illustration of the task studied in this paper and our proposed method. We aim to predict the distribution of responses from a set of listeners to noisy speech stimuli. We use Whisper, an ASR model, to predict a distribution over posible listeners'responses.}
    \label{fig:scheme}
\end{figure*}

\section{Problem}
\label{sec:problem}

We approach the problem of microscopic intelligibility prediction using the framework proposed by Marxer et al. \cite{marxer2015framework}. They consider a setup in which there are a set of tuples $(x, y, k)$, where $x \in {\mathbb{R}^T}$ is a noisy single-channel speech waveform of duration $T$, $y = [y_1, \dots, y_n]$ with $y_i \in \mathcal{V}$ for $i=1,\dots,n$ is a vector of $n$ unique responses perceived by a set of $m$ listeners from a finite vocabulary $\mathcal{V}$, and $k = [k_1, \dots, k_n]$ with $k_i \in \{1,\dots, m\}$ is a vector indicating the frequency of each response in $y$. They propose three tasks of increasing detail and difficulty to evaluate microscopic intelligibility models: confusion frequency prediction, confusion characterization, and full confusion prediction. The first task consists in predicting the probability of confusions occurring at the resolution of individual phonemes. In the second, the target is to predict what each phoneme will be confused with, i.e., predicting a probability distribution over possible phoneme substitutions. Finally, the last task is to predict the complete word perceived by the listeners, represented as a probability distribution over responses.

In this work, we focus on the most challenging task of full confusion prediction. Our model is therefore expected to predict a probability distribution over lexical responses conditioned on the speech waveform $P(y | x)$. As proposed by Marxer et al. \cite{marxer2015framework}, the performance of the model is evaluated as the likelihood of the observed data, namely the $(x, y, k)$ tuples, given the probabilities predicted by the model. The likelihood is the probability mass function of a multinomial distribution. Let $i$ index the tuples in a dataset of size $N$, $y_{i,j}$ and $k_{i,j}$ be the $j$-th response and its frequency in the $i$-th tuple, and let us denote $P(y_{i,j}|x_i)$ as $p_{i, j}$ for brevity. To avoid numerical underflow errors we use the log-likelihood. Then, the score $G$ that we seek to maximize is:

\begin{equation}
    G = \frac{1}{N} \sum_{i=1}^{N} \log{\left(\frac{m_i!}{k_{i, 1}!\dots k_{i, n_i}!} p_{i, j}^{k_{i, 1}} \dots p_{i, n_i}^{k_{i, n_i}}\right)}
    \label{eq:perf_logl}
\end{equation}

\section{Method}
\label{sec:method}

Our method is illustrated in Figure~\ref{fig:scheme}.
We use Whisper \cite{radford2023robust}, a recent ASR model trained on a 680,000 hours labeled speech corpus recorded in diverse conditions. Whisper outputs a probability distribution over word-level transcriptions given a speech input, therefore its application to lexical response prediction is straightforward. We chose Whisper mainly for two reasons: first, it is a state-of-the-art model whose accuracy and robustness approaches that of humans. We are interested on determining if its comparable performance to humans results in similar behavior under noisy speech stimuli. Second, its architecture is well suited for our task. Whisper consists of an encoder, which extracts contextual acoustic features $c$, and a decoder that predicts $P(y|c)$. Its encoder has been shown to produce useful features for macroscopic intelligibility prediction \cite{cuervo2024speech}. We would like to test if those representations are also useful for microscopic modeling. Having a separate decoder allows us to adapt it to our task without modifying the encoder. This is in contrast to other state-of-the-art encoder-only ASR systems in which the output is directly predicted from the encoder, e.g. \cite{baevski2020wav2vec}.

\subsection{Whisper architecture}
Whisper is an encoder-decoder transformer model \cite{vaswani17transformer}. The Mel-spectrogram of the speech waveform $x \in \mathbb{R}^T$ is fed to a strided convolutional neural network, which produces a sequence of frame representations $z \in \mathbb{R}^{T' \times d}$, where $d$ is the inner dimension of the model and $T' < T$. The encoder processes $z$ through a stack of transformer blocks with bidirectional self-attention, yielding a sequence of contextual features $c \in \mathbb{R}^{T' \times d}$. The decoder is a stack of transformer blocks with causal self-attention that predicts the probability over transcriptions conditioned through cross-attention on $c$. Whisper tokenizes transcriptions using subword units \cite{sennrich-etal-2016-neural}, therefore single words can be tokenized as a sequence of multiple tokens. For instance, the word \texttt{enjoyed} can be represented as the sequence of tokens $[$\texttt{enjoy}, \texttt{ed}$]$. Let $y$ be a single word tokenized as a sequence of tokens indexed by $t$, then the decoder outputs $P(y_t|y_{<t};c)$. By the chain rule of probability$P(y|c) = \prod_t p(y_t|y_{<t}; c)$. 

Whisper was trained on multiple tasks and languages, and it uses special tokens as a prefix in the decoder to inject information about the desired setup. We apply the model only for the task of ASR in English, therefore our prefix is in all cases the sequence of tokens \texttt{SOT} (start of transcription), \texttt{EN} (English), and \texttt{Transcribe} (ASR task).

\subsection{Transfer learning for lexical response prediction}

Whisper can be directly applied to lexical response prediction in a zero-shot fashion. Indeed, we can simply compute the likelihood of the listeners' responses predicted by the decoder. In this setup, we are essentially measuring the agreement between pre-trained Whisper and human speech perception.

We also fine-tune Whisper in two setups: to predict the top (most commonly reported) response, i.e. to match the majority of human listeners; and to model the distribution of listeners' responses. In the first case, we train the model to do ASR as in \cite{radford2023robust}, using the top response as target. In the latter, we train it to minimize the absolute difference between the predicted likelihood of the observed responses and their relative frequency:

\begin{equation}
    \mathcal{L} = \frac{1}{N} \sum_{i=1}^{N} \sum_{j=1}^n \left|\log(p_{i, j}) - \log\left({\frac{k_{i, j}}{\sum_{j=1}^n k_{i, j}}}\right)\right|
\end{equation}

\section{Experimental Setup}

\subsection{Dataset}

In our experiments, we used the publicly available English Consistent Confusion Corpus \cite{marxer2016corpus} (ECCC), a dataset created by gathering perceived responses from 15 listeners to common English words mixed with noise maskers of different types: stationary speech shaped noise (\texttt{SSN}); four-speaker speech babble (\texttt{BAB4}); and three-speaker babble modulated noise (\texttt{BMN3})). The corpus is composed of words-in-noise misperceived in the same way by at least 6 of the 15 listeners, resulting in over 3,000 consistent misperceptions. For each sample the corpus makes available the original utterance, the noise masker, and the list of listeners' responses.

We split the dataset in \texttt{train} (80\%), \texttt{dev} (10\%) and \texttt{test} (10\%) sets. Considering that the masker type induces different perceptual phenomena \cite{lecumberri16_interspeech}, and that other factors of variation such as speaker gender and identity are roughly balanced across maskers, the splits were made stratified by masker type. 

\begin{table}[t]
    \centering
    \caption{Performance on the \texttt{test} set for the different baselines (top and bottom) and transfer learning paradigms using \textsc{Whisper-Large-v3} (center).}
    \setlength{\tabcolsep}{2.5pt}
    \centering
    \setlength\extrarowheight{-3pt}
    \begin{tabular}{@{}lcccc@{}}
\toprule
\textbf{Model}  & \textbf{\makecell{Avg. log \\ likelihood}} & \textbf{\makecell{Top-1 \\ acc.}} & \textbf{\makecell{Avg. top-$n$ \\ coverage}} & \textbf{\makecell{Kendall \\ corr.}} \\ \midrule
Random          & -160                                       & 0.00                                & 0.00                                           &                0.00              \\
Multinomial     & -147                                       & 0.00                                & 0.00                                           & 0.11                    \\ 
Rnd. init. (pred. top)      & -175                                       & 0.00                              & 0.02                                         & 0.07                                 \\
Rnd. init. (pred. all)      & -176                                       & 0.00                              & 0.01                                         & 0.05                                 \\\midrule
Zero-shot       & -140                                       & 0.05                              & 0.25                                         & 0.28                                 \\
Fine-tuned (pred. top)   & -81                                        & 0.07                              & 0.39                                & 0.38                        \\
Fine-tuned (pred. all)  & \textbf{-46}                               & \textbf{0.13}                     & \textbf{0.48}                                & \textbf{0.42}                        \\ \midrule
Oracle          & -5                                         & 1                                 & 1                                            & 1                                    \\ \bottomrule
\end{tabular}
    \label{tab:performances}
\end{table}

\begin{table}[t]
    \centering
    \caption{Ablation of fine-tuned Whisper's modules. Models are trained to predict the distribution of listeners' responses.}
    \centering
    \setlength\extrarowheight{-3pt}
\begin{tabular}{@{}cccc@{}}
\toprule
\multicolumn{3}{c}{\textbf{Fine-tuned modules}}                                                                                               & \multirow{2}{*}{\begin{tabular}[c]{@{}c@{}}\textbf{Avg. log}\\ \textbf{likelihood}\end{tabular}} \\ \cmidrule(r){1-3}
\textbf{Decoder}    & \begin{tabular}[c]{@{}c@{}}\textbf{Encoder}\\ \textbf{Transformer}\end{tabular} & \begin{tabular}[c]{@{}c@{}}\textbf{Encoder}\\ \textbf{ConvNet} \end{tabular} &                                                                                \\ \midrule
\xmark & \xmark                                                        & \xmark                                                  & -140                                                                            \\
\checkmark & \xmark                                                        & \xmark                                                  & -59                                                                            \\
\checkmark & \checkmark                                                    & \xmark                                                  & -57                                                                            \\
\checkmark & \checkmark                                                    & \checkmark                                              & \textbf{-46}                                                                            \\ \bottomrule
\end{tabular}
    \label{tab:ablation}
\end{table}

\subsection{Evaluation}

As described in Section~\ref{sec:problem}, the criterion of performance for the task of lexical response prediction is the log-likelihood of the observed responses under the predictive model averaged across the dataset (Eq.~\ref{eq:perf_logl}). However, this metric is not easily interpretable, which motivated us to also report three other metrics: the accuracy of the model to predict the most common response as the most likely (top-1 accuracy), the average percentage of the $n$ responses present among the top-$n$ predictions of the model (average top-$n$ coverage), and the Kendall-Tau statistic \cite{kendall1945treatment}, which measures the correlations between the ground truth and predicted rankings of the observed responses.

For the top-1 and top-$n$ metrics, we compare phonemic representations, so that homophones are considered a match. We use the CMU pronunciation dictionary\footnote{\url{http://www.speech.cs.cmu.edu/cgi-bin/cmudict}} implemented in the \texttt{nltk} library \cite{bird2006nltk} to obtain phonemic representations.





For comparison, we report the same metrics for the naive baselines suggested by Marxer et al. \cite{marxer2015framework}: a random model (the frequency of a response is uniform across any possible word); a multinomial model empirically fitted to the responses independent from the speech stimuli; and a topline oracle model in which the predicted probabilities perfectly match the relative frequencies of listeners' responses. To assess the effect of transfer learning, we also consider a baseline in which we trained a model from scratch (i.e. with random initialization) on our task.

\subsection{Models and Training}

We used the official Whisper checkpoints available at {\small{\url{https://huggingface.co/openai}}}. For our fine-tuning experiments we used the Adam optimizer \cite{kingma15adam} with $\beta_1=0.9$, $\beta_2=0.999$, batch size of 16, and a learning rate schedule including a warm up phase. We did a hyperparameter search over the peak learning rate ($\{1\text{e-}3, 1\text{e-}4, 1\text{e-}5\}$), the fraction of training steps devoted to the warm up phase ($\{10\%, 50\%\}$), the learning rate schedule ($\{\text{linear}, \text{cosine}\}$) and the number of training epochs ($\{1, 4, 8, 12, 16\}$). As criterion to choose the best hyperparameters we used the log-likelihood of the data in the \texttt{dev} set. We did a full search using the \textsc{Whisper-Large-v3} model, from which the resulting best hyperparameters were: peak learning rate of 1e-5, warm up phase of 10\%, cosine learning rate decay, and 12 epochs of training. We used this setup for all the reported experiments. After training, we keep the checkpoint with the best performance on the \texttt{dev} set.



For all experiments we used a single NVIDIA A100 GPU with 80 GB of VRAM. A training epoch for \textsc{Whisper-Large-v3} takes 12 minutes and requires up to 37 GB of memory.


\begin{figure}[t]
    \centering
    \includegraphics[width=.65\linewidth]{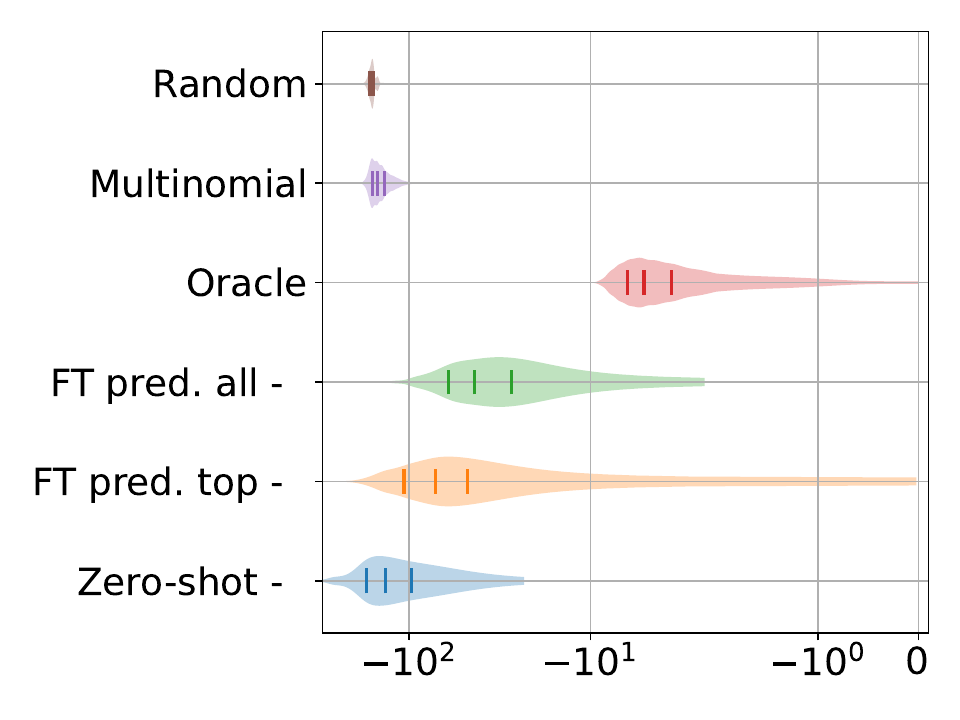}
    \caption{Distribution of responses' log-likelihoods predicted by different models. Vertical bars show quartiles.}
    \label{fig:likelihoods}
\end{figure}

\begin{table}[t]
    \centering
    \caption{Performance across model sizes when fine-tuning to predict the distribution of listeners' responses.}
    \centering
    \setlength\extrarowheight{-3pt}
    \begin{tabular}{@{}lcc@{}}
\toprule
\textbf{Pre-trained model}  & \textbf{\makecell{Parameters\\(millions)}} & \textbf{\makecell{Avg. log \\ likelihood}} \\ \midrule
\textsc{Whisper-Small}  &  244   & -53 \\
\textsc{Whisper-Medium} & 769 & -51 \\
\textsc{Whisper-Large-v3} & 1550 & \textbf{-46} \\ \bottomrule
\end{tabular}
    \label{tab:perf_sizes}
\end{table}

\begin{figure}[t]
    \centering
    \includegraphics[width=.49\linewidth]{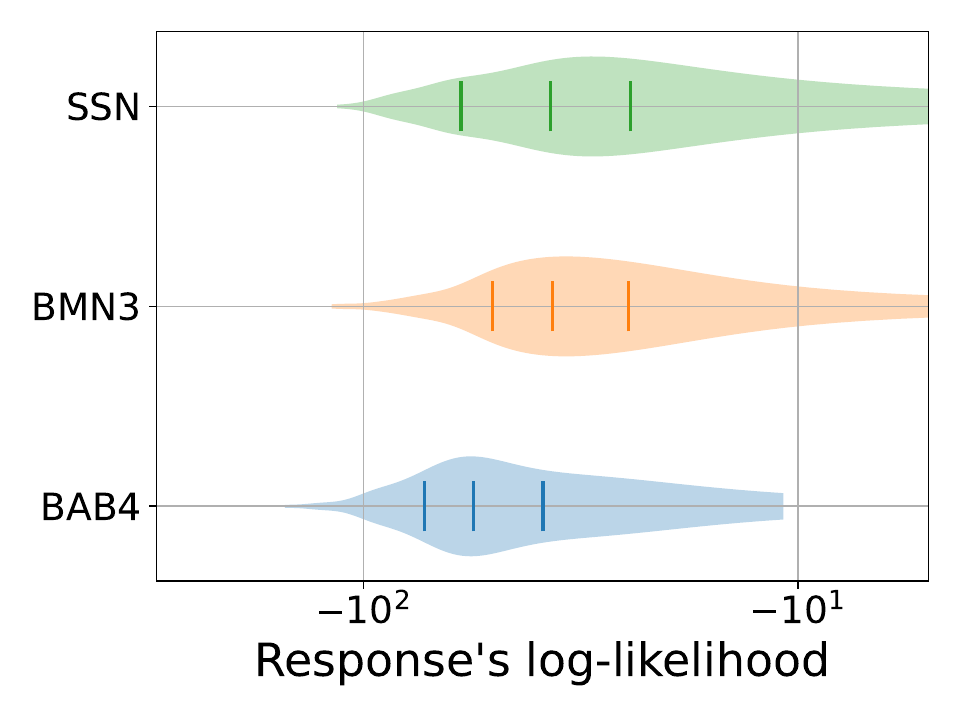}
    \includegraphics[width=.49\linewidth]{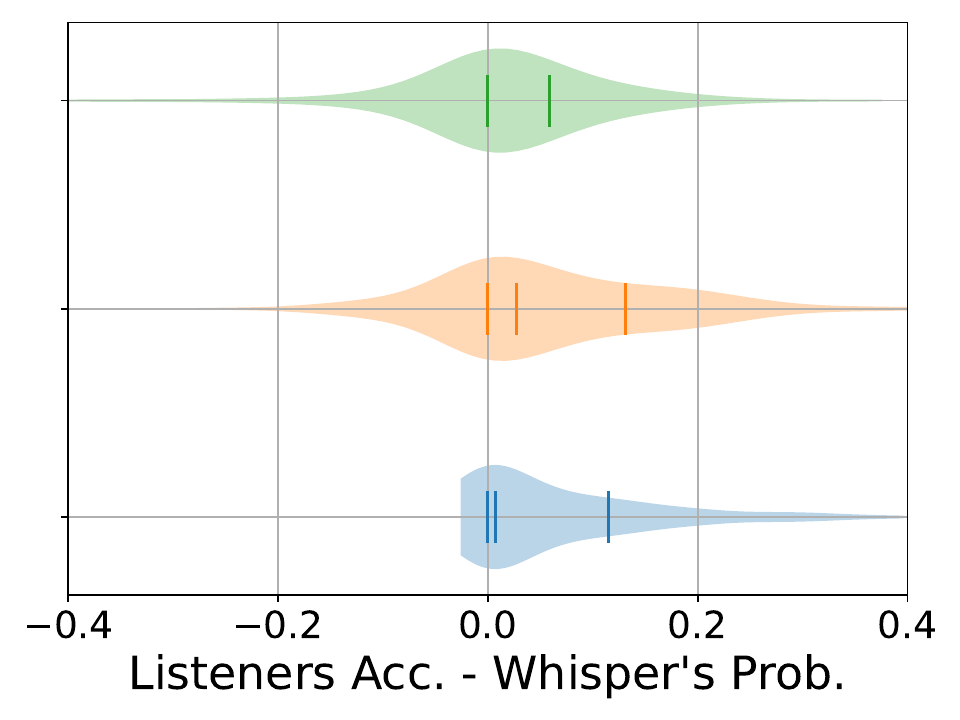}
    \caption{Variation of performance across masker types. (left) Distribution of responses' log-likelihoods. (right) Difference between listeners' relative frequencies and predicted probabilities for the spoken word. Vertical bars show quartiles.}
    \label{fig:masker_types}
\end{figure}

\section{Results}

\subsection{Effects of transfer learning}

On Table~\ref{tab:performances} and Figure~\ref{fig:likelihoods} we report the results obtained on the \texttt{test} set using the \textsc{Whisper-Large-v3} model. Our method outperformed all the considered baselines, even in a zero-shot setup. Models trained from scratch performed worse than the naive baselines, which shows the benefits of transfer learning on this task. Interestingly, the model fine-tuned to match the full distribution of responses ({\small{\texttt{FT pred. all}}}) performs better on top response prediction than the model trained specifically for it ({\small{\texttt{FT pred. top}}}). This shows that the extra supervision obtained from predicting the full distribution of responses leads to a better model of the majority's perception.

\subsection{Ablation of fine-tuning across modules}
As mentioned in Section~\ref{sec:method}, we wanted to determine the usefulness of each of the pre-trained Whisper modules for our task. In Table~\ref{tab:ablation} we show the performance when keeping some of the modules frozen. We saw little performance gain from fine-tuning the transformer in the encoder in addition to the decoder. However, there is a large relative improvement of over 40\% when fine-tuning the convolutional network in the encoder. This suggests that low-level acoustic features relevant for human perception are not being captured by Whisper's pre-training.

\subsection{Performance scaling with model size}

We also studied the scaling of performance with model size. We present the results in Table~\ref{tab:perf_sizes}. Empirical results show that larger models, which exhibit increased accuracy and robustness, are also better for human lexical response prediction.

\subsection{Performance by noise maskers}\label{chap:noise_maskers}

Motivated by the distinct effects on humans of different types of noise maskers in the ECCC \cite{lecumberri16_interspeech}, we studied their impact on performance. Figure~\ref{fig:masker_types} shows the obtained results. Our model has the most difficulty predicting human perception for speech corrupted with \texttt{BAB4} maskers. In order to assess if this mismatch was due to better or worse robustness relative to humans, we also evaluated the difference in accuracies between our model and humans, i.e. the difference between the probability of the correct response estimated by the listeners and that predicted by our model. For \texttt{BAB4}, the model was consistently worse than humans in perceiving the original spoken word (the distribution of differences is largely positive for \texttt{BAB4}). For the other masker types, in the majority of cases, humans also appear better than Whisper at perceiving the right word.

\section{Related work}
Previous works have used transfer learning from Whisper for macroscopic intelligibility prediction \cite{mogridge2024nonintrusive, zezario2023utilizing, cuervo2024speech}. Mogridge et al. \cite{mogridge2024nonintrusive} showed that Whisper's decoder representations outperformed those from the encoder, indicating that high level linguistic patterns heavily impact macroscopic intelligibility. 

As for microscopic intelligibility, Cuervo et al. \cite{cuervo23_interspeech} approached the task of phone-level prediction of misperception probability using transfer learning from wav2vec 2.0 \cite{baevski2020wav2vec}, a large scale self-supervised trained deep learning model. To the best of our knowledge, we are the first to use transfer learning from a large scale deep learning model for the micro-intelligibility task of lexical responses characterization.

\section{Discussion}

\subsection{On the implications of our results}

Our method significantly outperformed the baselines in all the considered metrics. This shows that Whisper does capture useful information for microscopic intelligibility prediction at the lexical level. Indeed, zero-shot experiments showed some degree of analogous behaviour between humans' and Whisper's perceptions of speech-in-noise (25\% of agreement on average). Moreover, we saw predictive performance increase with the size of the pre-trained Whisper model, similarly as to how accuracy and robustness scale \cite{radford2023robust}. Considering that the performance of deep learning models often scales predictably with the amount of compute (model size and amount of data) used for training \cite{hestness2017}, perhaps we can expect larger models, which are expected to be more accurate and robust, to increasingly converge with humans on speech perception.


Fine-tuning to match listeners' responses on the ECCC helped to significantly increase performance. Interestingly, we observed that fine-tuning to predict the full distribution of listeners' responses results in overall better performance than training to predict the mode of the distribution, even if one is only interested on predicting the mode. This shows that modeling fine-grained phenomena can bring benefits on coarser tasks, which cannot be obtained from training on the coarse task alone. Perhaps similar gains can be obtained on macroscopic intelligibility prediction by training on microscopic intelligibility prediction.

We obtained the most performance gains from fine-tuning the convolutional encoder, which suggests that the largest difference between Whisper's and human speech processing is at the low acoustic level. It could be interesting to see if a convolutional encoder fine-tuned for human speech perception would yield increased accuracy and/or robustness on ASR.

\subsection{On limitations}

We are limited mainly by the amount of data available for microscopic intelligibility prediction. We believe that part of the performance gap between our method and the topline can be due to the topline being a noisy, biased estimate of the real distribution of lexical responses that our model (by virtue of regularization) approximates. A larger sample of stimuli and listeners' responses is likely to meaningfully increase performance.

In terms of assessing the quality of our model, we are also limited by the lack of strong baselines that employ state-of-the-art methods for speech processing, against which we can compare our results. We hope this work will motivate other researchers to improve upon our work within a standardized evaluation framework, which will lead to the construction of a proper benchmark for microscopic intelligibility models.

\section{Conclusions}
We applied transfer learning methods from Whisper, a state-of-the-art automatic speech recognition system, for the task of predicting human lexical responses to speech-in-noise stimuli. We evaluated our model within a standardized framework in a zero-shot setup and when fine-tuned specifically for the task. Additionally, we studied the effects on performance of the granularity of fine-tuning, model size, fine-tuning of each pre-trained module, and different noise maskers. Results are promising, with our methods significantly outperforming naive baselines.



\section{Acknowledgements}

We are grateful to the French National Research Agency for their support through the ANR-20-CE23-0012-01 (MIM) grant, and the Institute of Convergence ILCB, supported by grants from France 2030 (ANR-16-CONV-0002) and the Excellence Initiative of Aix-Marseille University (A*MIDEX).


\bibliographystyle{IEEEtran}
\bibliography{mybib}

\end{document}